\documentclass[12pt,a4paper]{article}
\usepackage{a4wide}
\usepackage{latexsym}
\usepackage{epsf}
\usepackage{amssymb}
\def\[{\bigl[}
\def\]{\bigr]}
\def\({\bigl(}
\def\){\bigr)}

\def\o{\over}

\def\be{\begin{equation}}
\def\ee{\end{equation}}
\def\bea{\begin{eqnarray}}
\def\eea{\end{eqnarray}}
\def\nn{\nonumber}

\def\a{\alpha}
\def\b{\beta}
\def\g{\gamma}
\def\d{\delta}

\def\RR{\mathbb{R}}
\def\CC{\mathbb{C}}

\begin{document}
\begin{flushright}
IFT-UAM/CSIC-02-29\\
UFR-HEP/02-08\\ 
hep-th/0207160\\
\end{flushright}

\vspace{1cm}

\begin{center}

{\bf\Large On Non-Commutative Orbifolds of K3 Surfaces}

\vspace{.7cm}

{\bf A. Belhaj$^{1,}$\footnote{E-mail: {\tt ufrhep@fsr.ac.ma}},
J.J. Manjar\'\i n$^{2,}$\footnote{E-mail:
 {\tt juanjose.manjarin@uam.es}} and P. Resco$^{2,}$\footnote{E-mail: {\tt
 juanpedro.resco@uam.es}}}\\

\vspace{1cm}

$^1${\it  High Energy Physics Laboratory, Physics Department}
\\
 {Faculty of Sciences, Rabat, Morocco.}
\\[7pt]
$^2${\it
 Instituto de F\'{\i}sica Te\'orica, C-XVI,
  Universidad Aut\'onoma de Madrid} \\
  {E-28049-Madrid, Spain}\footnote{Unidad de Investigaci\'on Asociada
  al Centro de F\'{\i}sica Miguel Catal\'an (C.S.I.C.)}

\vskip 1.8cm

{\bf Abstract}
\end{center}
Using the algebraic geometry method of Berenstein and Leigh for the construction of the toroidal orbifold $ T^2\times T^2\times T^2 \over {Z_2 \times Z_2}$ with discrete torsion  and considering  local  K3 surfaces, we present  non-commutative aspects  of the orbifolds of product of  K3 surfaces. In this way,  the ordinary complex deformation of K3 can be identified with the  resolution of stringy singularities by non-commutative algebras using crossed products. We give representations and make some comments regarding the fractionation of branes. Illustrating examples are presented.
\newpage

%%%%%%%%%%%%%%%%%%%%%%%%%%%%%%%%%%%%%%%%%%%%%%%%%%%%%%%%%%%%%%%%%%%%%%%%%%
%%%%%%%%%%%%%%%%%%%%%%%%%%%%%%%%%%%%%%%%%%%%%%%%%%%%%%%%%%%%%%%%%%%%%%%%%%
 \section{Introduction}

It has been known for a long time that non-commutative (NC) geometry plays an interesting role in the context of string theory \cite{witten.1} and, more recently, in certain compactifications of the Matrix formulation of M-theory on NC torii \cite{co.do.sc.1}, which has opened new lines of research devoted, for example, to the study of solitons in connection with NC quantum field theories \cite{se.wi.1}.

In the context of superstring theory, NC geometry is involved whenever a $B$-field is turned on. For example,  in the study of $D(p-4)/Dp$ brane systems ($p>3$) where, in particular, one can consider the ADHM construction of the $D0/D4$ system \cite{gr.ne}, the NC version of the Nahm construction for monopoles \cite{gr.ne.1,bak,go.ma}, the determination of the vacuum field solutions of the Higgs branch of supersymmetric gauge theories with eight supercharges \cite{ah.be,be.sa1,be.sa2} or in the study of tachyon condensation using the so called GMS approach \cite{go.mi.st}.

However, most of the NC spaces considered in all these studies involve mainly NC $\RR_{\theta }^{d}$ \cite{go.mi.st}, NC torii T$_{\theta }^{d}$ \cite{sa.sa}, few cases of orbifolds of NC torii and some generalizations to NC higher dimensional cycles such as the NC Hizerbruch complex surface $F_{0}$ used in \cite{be.be.di.sa}.

Recently some efforts have been devoted to go beyond these geometric spaces. In particular, a special interest has been given to build NC Calabi-Yau (NCCY) manifolds containing the commutative ones as subalgebras and, in the case of orbifolds of Calabi-Yau (CY) threefolds, an explicit construction has been given by means of the so-called the NC algebraic geometric method \cite{be.le.1}. In that work, Berenstein and Leigh (BL) gave a realization of two NCCY 3-folds with discrete torsion:

\begin{enumerate}

\item  The toroidal orbifolds $\frac{T^6}{Z_2\times Z_2}$, where $T^6$ is viewed as the product of three elliptic curves as $T^2\times T^2\times T^2$. This construction involves non-commuting variables satisfying the 2-dimensional Clifford algebra.

\item The orbifold of the quintic in the $CP^4$ projective space,

\be
P_5(z_j)=z_1^{5}+z_2^{5}+...+z_{5}^{5} +\lambda \prod_{i=1}^{5}z_i= 0,
\ee

\noindent by the $ Z^{3}_{5}$ discrete torsion symmetry group. The quintic algebra ${\cal{A}}_{\theta}(5)$ reads as:

\bea
z_{1}z_{2}=\alpha z_{2}z_{1},\qquad z_{1}z_{3}=\alpha^{-1}\beta z_{3}z_{1},\nn\\
z_{1}z_{4}=\beta^{-1}z_{4}z_{1},\qquad z_{2}z_{3}=\alpha\gamma z_{3}z_{2},\nn\\
z_{2}z_{4}=\gamma ^{-1}z_{4}z_{2},\qquad z_{3}z_{4}=\beta\gamma z_{4}z_{3}, \\
z_{i}z_{5}=z_{5}z_{i},\qquad i =1,2,3,4,\nn
\eea

\noindent where $\a$, $\b$ and $\g$ are fifth roots of the unity generating the ${Z}_{5}^{3}$\ discrete group and where the $z_{i}$'s are now the generators of the quintic algebra.

\end{enumerate}

In this context, thinking of D-branes as coherent sheaves with support on a NC subvariety, they also explained the fractionation of branes by using a limit where the rank of the sheaf could jump at the singularity, leading to reducible matrix representations of the algebra.

Such formulation has been extended to higher dimensional orbifolds, understood as homogeneous hypersurfaces $P_{n+2}(z_1,z_2,...,z_{n+2})$ in ${\bf CP^{n+1}}$ with some discrete group of isometries $\bf Z^{n(n+1)/2}_{n+2}$ \cite{be.sa3}. In all these works, the CY algebra has a typical form which reminds quantum groups and the Yang-Baxter equations \cite{lo.sc.we,we}:

\be
z_{i }z_{j}={\cal{R}}_{ij }{}^{\alpha \beta }z_{\alpha }z_{\beta },
\ee

\noindent where the four rank tensor ${\cal{R}}_{ij }{}^{\alpha \beta }$ was determined by the discrete torsion and the CY conditions. As we will see in this work, these ${\cal{R}}_{ij}{}^{\alpha \beta }$ can take the following form

\be
{\cal{R}}_{ij }{}^{\a\b}=\d_{i }{}^{\b}\d_{j}{}^{\a}w^{{\ell_{ij}}},
\ee

\noindent where $w$ is an element of the discrete group $G$, which leaves invariant the CY algebraic equation, and $\ell_{ij}$ is an antisymmetric matrix satisfying the identity $\sum_{i}\ell_{ij}=0$ wich can be interpreted as the CY condition.

This  analysis can also be adapted for lower dimensional CY manifolds \cite{ki.lr1}. In particular, we are interested in the case of the $K3$ surface. This is a very special surface because it is the only two-dimensional CY manifold. It can be represented in different ways depending on which property one is willing to
study.

It can be easily seen to be related to the superconformal model corresponding to the polynomial constraint in $WCP^3_{1,1,1,1}$ plus deformations \cite{gr.va.wa.1,va.wi.1} and so can be seen as a complex surface in this space. This representation  makes very clear the complex structure of the surface. Another description is a local one in terms of the ADE classification of singularities near the singular loci of the orbifold $T^4/Z_2$. A third description is in terms of an elliptic fibration, which means that locally the surface could be seen as a two torus times a complex plane.

All along the paper we will be dealing with the two first descriptions, although the last one could be used to find a proper interpretation of the results we will find, as will be explained in the conclusions.

The aim of this work is to extend the results found in the case of the orbifold $T^6\o Z_2\times Z_2$ to higher CY manifolds in terms of product of K3 surfaces and, as we are considering $Z_2\times Z_2$ orbifolds, which have $H^2(Z_2\times Z_2,U(1))\simeq Z_2$, we can include the effect of discrete torsion.

It is known that when the discrete torsion is considered, the twisted sector modes are in $H^{2,1}$ and so act in the deformation of the complex structure of the orbifold \cite{va.wi.1}. However, there are not enough deformations available to resolve the singularities, because the discrete torsion is supported at  them. In these cases, the only known way to resolve the singularities of the space is via NC geometry.

The outline of the paper is as follows. In section \ref{ade} we review the basic facts of the construction of the K3 surfaces in terms of the ADE classification of singularities and study the deformations which can be made to the equations which define them. In section \ref{alg} we will study how to construct the NC algebra associated to the orbifolds of CY manifolds. In section \ref{orb} we specialize to the case of orbifolding the product of three K3 and construct the realization of the associated algebra. In section \ref{morb} we extend the study to higher dimensional cases. We finish in section \ref{cd} with a discussion and some conclusions.

%%%%%%%%%%%%%%%%%%%%%%%%%%%%%%%%%%%%%%%%%%%%%%%%%%%%%%%%%%%%%%%%%%%%%%%%%%%%%%%
%%%%%%%%%%%%%%%%%%%%%%%%%%%%%%%%%%%%%%%%%%%%%%%%%%%%%%%%%%%%%%%%%%%%%%%%%%%%%%%
\section{ K3 surfaces, with ADE singularities, in string theory compactifications}\label{ade}

 In this section we give certain essential aspects of K3 surfaces as well
 as methods for the resolution of ADE singularities. This study is based on the results
  of the geometric engineering of $D=4$ $N=2$ quantum field theory embedded in superstring
   theory compatifications \cite{ka.ma.va,ma,be.fa.sa}.

Roughly speaking, K3 is a two complex dimensional compact K\"ahler
CY manifold with SU(2) holonomy group. It has many types of
realizations, the simplest one is to consider the orbifold
$T^4\over G$, where $T^4$ is defined by the following complex
identification equations

\bea
z_j&\equiv& z_j+1,\nn\\
z_j&\equiv& z_j+i,\quad j=1,2,
\eea

\noindent and where $G$ is a discrete subgroup of SU(2). For
instance, if we consider $G=Z_2$, the K3 surface is obtained by
imposing a  extra constraint equations on $T^4$, namely

\be
z_j\equiv -z_j,\quad j=1,2.
\ee

This symmetry has sixteen singular fixed points. Near such points ($z_1,z_2)\equiv(-z_1,-z_2)$, the K3 surface looks like $\CC^2\o {Z}_2$ and can be determined algebraically in terms of the ${ Z}_2$ invariant coordinates on $\CC^2$, which are given by

\bea
\label{xyz}
x&=&z_1^2\nn\\
y&=&z_2^2\\
z&=&z_1z_2,\nn
\eea

\noindent and give a map form $\CC^2\o { Z}_2$  to $\CC^3$.

Locally, K3 can be viewed as a hypersurface in $\CC^3$ defined by

\be
\label{a1}
z^2=xy
\ee

\noindent The equation (\ref{a1}), which is known by $A_1$ singularity, can be
 extended to the so called $A_{n-1}$ singularity having the following form

\be
\label{an}
A_{n-1}:z^{n}= xy.
\ee

Other singularities of local K3 surfaces are classified by the
following equations

\be
\matrix{D_n: & x^2+y^2z+z^{n-1}=0,\cr \\
                 E_6: & x^2+y^3+z^4=0,\cr\\
                 E_7: & x^2+y^3+yz^3=0,\cr\\
                 E_8: & x^2+y^3+z^5=0.}
\ee

Basically there are two ways for smoothing out the ADE singularities, either
by deforming its K\"ahler or its complex structure. For later use we shall
focus our attention on the resolution of the $A_{n-1}$ singularity, where the
complex deformation deals with the left hand side of equation (\ref{an}),
while the K\"ahler deformation, which consists in blowing up the singular
point with the help of $(n-1)$ intersecting real $S^2$, treats the right hand
side of equation (\ref{an}). In the case of K3 seen as an $A_{1}$ singularity
these two operations are related, because the $A_1$ singularity can be seen as
a vanishing 2-sphere, so either deforming the complex structure or making a
blow-up consists in giving finite volume to it.

This method has a very nice interpretation in terms of the toric geometry realization
of local K3 surfaces, where the Mori vectors are intimately related to the $A_{n-1}$
Cartan subalgebra charges of the gauge symmetry involved in the geometric engineering method. Moreover, the corresponding toric graph looks similar to the $A_{n-1}$ Dynkin diagram.

Since this method of doing is mirror to the complex deformation
and, for latter use, we will only  give the complex deformation of
the $A_{n-1}$ singularity. Indeed,  equation (\ref{an}) admits a
discrete ${ Z}_n$ symmetry acting as follows

\be
\matrix{z\rightarrow wz, & w^n=1 \cr\\
x\rightarrow x, & \cr\\
y\rightarrow y, &}
\ee

\noindent leaving $x$ and $y$ invariants.  The deformation of the
complex structure of the $A_{n-1}$ singularity introduces extra
terms breaking the ${ Z}_n$ symmetry as follows

\be
xy=z^n+P(z).
\ee

In this equation, the extra polynomial is given by

\be
\label{poly}
P(z)=\sum \limits_{i=1}^{n-1}a_iz^{n-i-1},
\ee

\noindent where the  $a_i$'s are complex parameters carrying the complex deformation of the $A_{n-1}$ singularity. Their number is $(n-1)$ which is the rank of the $A_{n-1}$ Lie algebra.  These results have been used in many directions in string theory and F-theory compactifications, in particular, in the study  of the quantum field theory using the geometric engineering method.

It should be interesting to note the following points for the
$A_{n-1}$ geometry

\begin{enumerate}

\item Since K3 is a self mirror, equation (\ref{poly}) means that each monomial $z^k$ is associated to a divisor of K3 explaining the monomial/divisor map involved in the mirror symmetry application in the toric geometry framework.

\item The complex deformation acts only on the $z$ variable, by introducing terms breaking the $Z_n$ symmetry. The restoration of this symmetry leads to a limit where K3 develops the $A_{n-1}$ singularity.

\item  The complex deformation of $A_{n-1}$ singularity is similar to the resolution
 of stringy singularities by a NC algebra involved in the study of the orbifold
  $C^2\o Z_2$ using the crossed product algebra \cite{be.le.2}. Indeed, identifying the role of
  the ${ Z}_n$ symmetry involved in the complex deformation with the ${ Z}_n$ discrete
   torsion of the crossed product of $C^2\o { Z}_2$, one  can identify the complex deformation
    and the resolution of the stringy singularity of the orbifold $C^2\o Z_2$.

This link can be understood by the fact the center of the algebras, being the singular geometry, is invariant under the ${ Z}_n$ symmetry corresponding to $P(z)=0$ in the commutative deformation. Taking into account this fact, one can see that the terms of the deformation, in the NC sense, must not be in the center of the algebra. By this argument, one can see that the complex deformation of $C^2\o Z_2$, in the commutative sense, is similar to the stringy singularities by NC algebra involved in the study of the orbifold $C^2\o Z_2$ using NC algebraic geometry method and the crossed product algebra.

\end{enumerate}

%%%%%%%%%%%%%%%%%%%%%%%%%%%%%%%%%%%%%%%%%%%%%%%%%%%%%%%%%%%%%%%%%%%%%%%%%%%%%%%
%%%%%%%%%%%%%%%%%%%%%%%%%%%%%%%%%%%%%%%%%%%%%%%%%%%%%%%%%%%%%%%%%%%%%%%%%%%%%%%
 \section{NC algebraic geometry method}\label{alg}

In this section, we will briefily review the NC algebraic geometry approach, introduced
 first in \cite{be.le.1}, for treating the NC aspects of orbifolds of the CY manifolds. In this method, the (singular) orbifold with discrete torsion can be viewed as a NC algebra. In other words, the algebraic realization of a commutative orbifold space with discrete torsion has a nice interpretation using NC algebra.

In this method ones proceeds following the next steps. Firstly, one takes a
$d$-dimensional (singular) complex CY manifold $M^d$ defined by an equation of the form

\bea
f_j(u_i)=0, \quad i-j=d,
\eea

\noindent where the $u_i$ are complex local coordinates. One looks for a discrete symmetry $G$

\bea
G:\quad u_i\to gu_i,\quad g\in G,
\eea

\noindent leaving $f_j(u_i)$ invariant

\bea
G:\quad f_j(u_i)\to f_j(u_i),
\eea

\noindent and preserving the CY condition. After that, one considers the orbifold $M^d\over G$ which is constructed by identifying the points which are in the same orbit under the action of the group, i.e., $u_i\to gu_i$. The resulting space is smooth every where, except at the fixed points, which are invariant under non trivial group elements.

Following \cite{be.le.1,be.sa3,ki.lr1,be.le.2,be.je.le.1,sa} and using the discrete symmetry group $G$, one can build the NC extensions of the above orbifold, $\left({{{\cal{M}}^d \o G}}\right)_{nc}$. This procedure may be summarized as follows: the NC extension of this orbifold is obtained, as usual, by extending the commutative algebra ${\cal{A}}_{c}$ of functions on $M^d\over G$ to a NC one ${\cal{A}}_{nc}\sim {\left({{\cal{M}}^d \o G}\right)}_{nc}$. In this algebra, the coordinate functions $u_{i}$ on the deformed geometry will obey the following constraint equations

\be
u_iu_j=\theta_{ij}u_ju_i,
\ee

\noindent where $\theta_{ij}$ are the NC parameters constrained by

\bea
\theta_{ij}\in G,\nn\\ 
\theta_{ij}\theta_{ji}=1.
\eea 

As we will see, the solution of these equations can take the following form

\be
\theta_{ij }=g^{{\ell_{ij}}},
\ee

\noindent  where $g$  are the generators of $G$ and $\ell_{ij}$ is an antisymmetric tensor. An explicit solution is obtained with the help of extra constraints on the $\theta_{ij}$'s which can be easily specified once we know the elements of the center of the NC version of the orbifold, ${\cal Z}({\cal{A}}_{nc})$.

The elements of ${\cal Z}({\cal{A}}_{nc})$, which yield the commutative algebra, are the quantities invariant under the action of $G$. In this way, the algebraic geometry of ${\cal Z}({\cal{A}}_{\theta})$ is identified with the algebraic realization (\ref{poly}), which may be singular, while the algebraic geometry of the NC algebra will resolve the singularities. In other words, the commutative singularity can be deformed in a NC algebraic realization sense.

Since the deformation part is not invariant under $G$, one may say that this part resolving the singularity must be in ${{\cal{A}}_{\theta}\over  {\cal Z}({\cal{A}}_{\theta})}$ which may be a NC subspace algebra of ${\cal{A}}_{nc}$. By this argument, we think that the same feature appears in the ordinary complex deformation of $A_{n-1}$ singularity of $K3$ surfaces where the extra terms solving the singularity are not invariant under the ${ Z}_n$ symmetry.

This important link between the complex deformations and the resolution of stringy singularities by the NC algebras push us to think about the extension of the result of BL concerning the orbifold of the torus $ T^6\o Z_2\times Z_2$ in terms of K3 surfaces using NC algebraic geometry method.
Before doing this, let us first recall the BL work for $ T^6\o Z_2\times
Z_2$. In this work, $T^6$ is viewed as the product of three elliptic curves as
$T^2\times T^2\times T^2$, each given in the Weierstrass form

\be
y_i^2 = x_i(x_i-1)(x_i-a_i), \quad i=1,2,3,
\ee

\noindent  for $i=1,2,3$, with a point added at infinity. The later can be brought to a finite point by a change of variables

\bea
\label{tos}
y_i\to y'_i={y_i\o x_i^2}\nn \\
x_i\to x'_i={1\o x_i}
\eea

The ${ Z_2\times Z_2}$ discrete symmetry acts by $y_i\to \pm y_i$ and $x_i\to x_i$ so that the
 holomorphic three form $dy_1\wedge dy_2 \wedge dy_3$ is invariant under the orbifold action satisfying the CY condition. After introducing the discrete torsion, the constraints of the NC reads

\be
\matrix{y_i y_j = - y_j y_i, & \hbox{for } i\neq j\cr \\
                 x_i x_j = x_j x_i,   & \hbox{for } i,j=1,2,3,\cr\\
                 x_i y_j = y_j x_i.   &}
\ee

\noindent and can be solved by

\bea
y_i&=&a_i\sigma_i,\nn\\
x_i&=&b_i I_2.
\eea

By this approach, the orbifold $T^6\over{Z_2}\times { Z_2}$ with torsion defines a NCCY threefold, where the NC is carried by the discrete torsion phases and having a remarkable interpretation in terms of closed string states. On the fixed planes, the branes fractionate and local deformations are no more trivial. In what follows, we want to extend this result to higher dimensional CY manifolds. In particular we will consider CY's realized as orbifolds of $K3$ surfaces with discrete torsion. In other words, instead of having products of the $T^2$ elliptic curves, we will have products of K3 surfaces.

%%%%%%%%%%%%%%%%%%%%%%%%%%%%%%%%%%%%%%%%%%%%%%%%%%%%%%%%%%%%%%%%%%%%%%%%%
%%%%%%%%%%%%%%%%%%%%%%%%%%%%%%%%%%%%%%%%%%%%%%%%%%%%%%%%%%%%%%%%%%%%%%%%%
\section{NC orbifolds of the K3 surfaces}\label{orb}

 In this section, we  start by consider a general $K3$ \cite{ki.lr2}. The latter  are given by the following general form and with a point added at infinity

\be
\label{fxy}
 z^2 = f(x,y),
 \ee

\noindent where $f$ is obtained from a homogeneous function $F$ with total degree 6 in complex variables $u,v,w$ as follows

\be
\label{f6}
F(u,v,w)= F_6(u,v,w).
\ee

Note  that a special form which has been used in \cite{ki.lr2} for studying $N$-point deformation of  algebraic K3 surfaces is given by

\be
\label{fuvw}
F(u,v,w)= u^2v^3w+u^4v^2.
\ee

However, in order to connect the algebraic geometry (\ref{fxy}) to ones described in section \ref{ade}, we will take here  a special form of (\ref{f6}) as follows

\be
\label{fu4vw} 
F(u,v,w)= u^4vw.
\ee

By this form, it is not difficult to see that (\ref{fxy}) leads to the algebraic equation describing the $A_1$ singularity of $K3$ surfaces. Indeed, dividing  equation (\ref{fu4vw}) by $u^6$, one obtains

\be
{F(u,v,w)\o u^6} =\left({v\o u}\right)\left({w\o u}\right),
\ee

\noindent and so $f$ is given by

\be
f(x,y)=xy,
\ee

\noindent where

\bea
x&=& {v\o u}, \nn\\
y&=&{w\o u}.
\eea

In this case, (\ref{fxy})  looks like as  the  ALE space with  $A_1$ singularity given by (\ref{xyz}), and the analogue of the equations (\ref{tos}) reads now as

\bea\label{change}
z&\to  &z'={z\o x^3},\nn\\
x&\to &  x'={y\o x}, \\
y&\to  &y'={1\o x}\nn.
\eea

 By these equations, now we are in position to extend the results of the orbifold $T^6\over\bf{Z_2}\times {Z_2}$ with discrete torsion. To start, we consider the following geometric realization of $\frac{K3^{\otimes3}}{G}$, that is, $K3^{\otimes3}$ is represented by the product of three  K3  as follows

\be
\label{cosa}
z_i^{2}=x_iy_i,\quad i=1,2,3,
\ee

\noindent with an orbifold group $G$ specified later on. A priori there are different symmetries leaving these equations invariant, but in order to keep the same analysis of \cite{be.le.1}, we will take $G$ as ${ Z_2 ^2}$ acting only on the $z_i$variables as follows

\bea
\label{vars}
z_i &\rightarrow &\pm z_i,\nn\\
x_i &\rightarrow & x_i, \\ y_i
&\rightarrow & y_i .\nn
\eea

The reason behind choosing this symmetry is that the complex deformation of K3 surfaces acts only  on the  each  $z$ variable of K3 surfaces. The CY condition of this orbifold requiresthat the holomorphic six form

\be
\Omega_6=dz_1\wedge dz_2 \wedge dz_3 \wedge \frac{dx_1}{y_1}\wedge \frac{dx_2}{y_2}\wedge \frac{dx_3}{y_3},
\ee

\noindent should be invariant under (\ref{vars}). Furthermore, since the $Z_2\times Z_2$ symmetry acts only on $z_i$, it follows that the invariance $\Omega_6$ is reduced to  the invariance of  $dz_1\wedge dz_2\wedge dz_3$. Having introduced these data, now we would like to introduce the discrete torsion. The orbifold $K3\times K3\times K3 \o { Z_2^{2} }$ with discrete torsion can be viewed as a NC hyper-Kahler CY manifold, where the ${ Z_2\times Z_2}$ invariant terms are elements of the center of the algebra.  Using this feature and the CY condition, the NC version of the orbifold $K3^{3}\o { Z_2^2}$ is obtained by taking the coordinates functions as follows

\bea
\label{alg1}
z_{i}z_{j} &=&-z_{j}z_{i},\nn\\
z_{i}x_{j} &=& x_{j}z_{i},\nn\\
y_{i}z_{j} &=& z_{j}y_{i},\\
x_{i}x_{j} &=& x_{j}x_{i},\nn\\
y_{i}y_{j} &=& y_{j}y_{i},\nn\\
y_{i}x_{j} &=& x_{j}y_{i},\nn
\eea

\noindent with

\be
\label{alg2}
\matrix{z_{i }^2z_{j } & = & z_{j }z_{i }^2,\cr\\
\[ z_{i},\prod_{i=1}^{3}z_i\] & = & 0,}
\ee

\noindent which means that the $x_{i}$, $y_{i}$, $z^2_{i}$ and $\prod\limits_i  z_{i}$ are all in the center of the algebra. To find the points of the NC geometry, the algebra (\ref{alg1}-\ref{alg2}) can be represented in terms of the Pauli matrices as follows

\bea
z_i=a_i\sigma_i,\nn\\
x_i=b_i I_2,\\
y_i=c_i I_2,\nn
\eea

\noindent where the $a_i, b_i,c_i$ are complex scalars and $I_2$ is the two dimensional identity matrix. Since  the algebra (\ref{alg1}-\ref{alg2}) is very similar to the one describing the NC version of the orbifold torus, it follow that one should have the same interpretation in terms of resolution of singularities and reducibility of representations.

We can make the following remarks about the analysis made: the first one, which will be given in this section, is that we may find a Clifford algebra, in particular the Dirac algebra involved in the quantum field theory on Euclidean space. Another point is to consider the higher order of the discrete symmetries appearing in the  geometry of ALE space, which will be treated in the next section.

Before going ahead, let us recall some useful properties of the Dirac algebra. The later, which is involved in the study of fermions, is given by

\be
\{\gamma_i,\gamma_j\}=\delta_{ij},
\ee

\noindent where $\gamma_i$ are complex matrices satisfying

\bea\label{dirac}
\gamma^{\dag}_0&=&\gamma_0,\nn\\
\gamma^{\dag}_i&=&-\gamma_i,\\
\gamma^{\dag}_5&=&\gamma_5.\nn
\eea

Note that the minus sign in (\ref{dirac}) can be absorbed by transforming $\gamma_i\rightarrow i\gamma_i$, giving hermitian Dirac matrices which will be useful for discussing the brane fractionation in this context.

As an illustrating application, we can consider the product of five K3 surfaces with ${ Z_2^5}$ discrete symmetry. The later acts on  $ z_i$, $x_i$'s and $y_i$'s as:

\bea
z_{i } &\rightarrow &z_{i }^{\prime }=z_{i }\omega ^{q^a_{i}}\quad a=1,2,3,4\nn\\ 
x_{i } &\rightarrow &x_{i }^{\prime }=x_{i},\\ 
y_{i } &\rightarrow &y_{i }^{\prime }=y_{i }, \nn
\eea

\noindent where  $\omega =\pm 1$ and $q_i^a$ are integer vectors satisfying the CY condition

\be
\sum_{i =1}^{5}q_{i }^{a}=0,\quad \mbox{ mod 2};\quad a=1,...,4.
\ee

Using the constraints on the $\theta$ parameters and the CY condition, one  can write

\bea
\theta _{ij } &=&(-1)^{\ell_{ij }}, \nn\\ 
\sum_{i =1}^{5}\ell_{ij} &=&0,\quad \mbox{mod 2},
\eea

\noindent where $\ell _{ij }$ is an antisymmetric matrix of the following form

\be
\ell_{ij }=\Omega _{ab}q_{i }^{a}q_{j }^{b};
\ee

\noindent where $\Omega _{ab}=-\Omega _{ba},$ and $\Omega _{ab}=1$ for $a<b$.

Now, if we take $\ell_{ij }=1$ that is

\be
\theta_{ij}=-1\quad  \forall \;i\neq j
 \ee

\noindent the  NC algebra reduces to 

\be
\label{su2} 
z_iz_j=-z_jz_i, \quad \hbox{for } i\neq j, \quad \hbox{for } \quad i,j=1,\ldots,5,
\ee

\noindent with all others commutations relations. Using the Dirac matrices, a four dimensional realization of the algebra (\ref{su2}) can be written as follows.

\bea
z_i=a_i\gamma_i\nn\\
x_i=b_i I_4\\
y_i=c_i I_4\nn
\eea

\noindent where now $\gamma_i$ are given by

\be
\gamma_1=\pmatrix{ I_2 & 0\cr 0 &-I_2},\quad
\gamma_{i+1}=\pmatrix{ 0 &\sigma_i\cr -\sigma_i & 0},i=1,2,3\quad
\gamma_5=-i\pmatrix{ 0&I_2\cr I_2&0},
\ee

\noindent and where the $\sigma_i$ are the Pauli matrices. At fixed locus, this representation becomes reducible as four out of the five variables $z_i$ act by zero. Thus we get four distinct NC points, and so there four different irreducibles representations corresponding to the four eigenvalues of the non zero $z_i$.

%%%%%%%%%%%%%%%%%%%%%%%%%%%%%%%%%%%%%%%%%%%%%%%%%%%%%%%%%
%%%%%%%%%%%%%%%%%%%%%%%%%%%%%%%%%%%%%%%%%%%%%%%%%%%%%%%%%
\section{More on the Orbifolds of K3 surfaces}\label{morb}

As we have mentioned, the above geometry can be extended to ones with  higher dimensional discrete symmetries. In this case, the  analogue of  equation of (\ref{cosa}) is

\begin{equation}\label{cosa2}
z_{i }^{n}=x_{i}y_{i},\quad i =1,...,m,
\end{equation}

\noindent where $m$  is an integer, which will be fixed later on. As in the previous examples, equations (\ref{cosa2}) have a $ Z_n^{m-1}$ discrete group symmetry acting  on the variables as follows

\begin{eqnarray}
z_{i} &\rightarrow &z_{i }^{\prime }=\omega ^{q^a_{i }}z_{i }\quad a=1,...,m-1, \nn \\ 
x_{i} &\rightarrow &x_{i}^{\prime }=x_{i },\\ 
y_{i} &\rightarrow &y_{i}^{\prime }=y_{i },\nn
\end{eqnarray}

\noindent so that  $dz_1\wedge dz_2\wedge\ldots\wedge d z_m$  is invariant. This satisfies the CY condition on the quotient space. In equation (\ref{su2}), $\omega $ is an element of the discrete group $ Z_n^{m-1}$ and where $q^a_{i }$ are integers  satisfying the  following  condition $\sum_{i =1}^{n+1}q^a_{i}=0,\quad mod\; m $ which is also interpreted as the CY condition. Using the previous analysis, the NC version of the orbifold $K3\otimes^m\o {Z_2^{m -1}}$ is obtained by substituting the usual commutative algebra of the functions by the NC one. In this way, the coordinate functions $x_{i }$, $y_{i }$ and $z_{i }$ on the deformed NC manifold obey the following identities

\bea\label{pepita}
z_{i }z_{j }&=&\theta _{i j }z_{j }z_{i },\nn\\
z_{i }x_{j }&=&x_{j }z_{i },\nn\\
y_{i }z_{j }&=&z_{j }y_{i },\\
x_{i }x_{j }&=&x_{j }x_{i },\nn\\
y_{i }y_{j }&=&y_{j }y_{i },\nn\\
y_{i }x_{j }&=&x_{j }y_{i },\nn
\eea

\noindent  with

\be
\matrix{ z_{i }^nz_{j } & = & z_{j }z_{i }^n,\cr 
                  \[ z_{i},\prod_{i=1}^{m}z_i\]& = & 0},
\ee

\noindent which means that $z_{i }^n$ and  $\prod_{i=1}^{m}z_i$ belong  to the center of the NC algebra. Using  all these identities, one can easily see that the $\theta_{ij }$ parameters must satisfy the following constraint equations

\bea
\label{xi}
\theta _{ij}^{n} &=&1, \\ 
\prod_{i=1}^{m}\theta _{ij} &=&1,\qquad \forall i, \\ 
\theta _{ij}\theta _{ji} &=&1.
\eea

These constraints can be solved as follows: First, equations (\ref{xi}) show that

\be
\theta _{ij }=\omega ^{\ell _{ij }}, \quad \omega= exp
\frac{2i\pi}{n}
\ee

\noindent where $\ell _{ij  }$ is a $m\times m $  matrix. Second,  putting this equation back into (\ref{xi}), one finds that   $\ell_{ij }$  must satisfy

\be
\matrix{ \ell_{ij } =-\ell_{ji}, & \cr
                  \sum_{i =1}^{m}\ell_{ij } =0, &\mbox {mod n.}}
\ee

Next  we will build the irreducible representations of the NCCY algebra for a regular representation. Then we will give the representation for the fixed points( where becomes reducible). It turns out that the $d$  dimension of the  finite matrix representations of the orbifolds geometry algebra is a multiple of $n$.  To see this property it is enough to take the determinant of both sides of NC variables namely

\be
det\;(z_{i}z_{j}) =({\theta _{ij}})^d det\; (z_{j}z_{i})= det\;(z_{j}z_{i})
\ee

\noindent which constraint the dimension $d$ of the representation to be such that:

\be
\theta_{ij}^{d}=1.
\ee

\noindent Using the identity (\ref{xi}), one discovers that $d$ is a multiple of $n$.

We  return to equation (\ref{cosa2}), the change of variables (\ref{change}) takes now the following form

\bea 
z_i&\to &{z_i\o x_i^{6/n}}\nn\\
x_i&\to& x'_i={y_i\o x_i} \\
y_i&\to& y'_i={1\o x_i}.\nn
\eea

If we require that $6/n$ must be integer, therefore one has only  $n=2,3,6$.

\begin{enumerate}

\item {\bf Case of  $n=2$}

we get the geometry related to $A_1$ singularity, described in section 4.

\item {\bf Case of $n=3$}

Instead of being general, we give a concrete example corresponding to  $m=3$. In this case the equation (\ref{cosa2}) reduces to

\begin{equation}
z_{i }^{3}=x_{i}y_{i},\quad i =1,2,3.
\end{equation}

\noindent being the $A_2$ singularity. Of course  the NC version of this geometry is obtained from  the one given in (\ref{pepita}). In this case, the equations (\ref{xi}) can be solved as follows

\be
\theta_{ij}=\omega^{\ell_{ij}},
\ee

\noindent where $\omega$ is a phase so that $\omega^3=1$ and $\ell_{ij}$ is $3\times 3$ antisymmetric matrix

\be
\ell_{ij}=\left(\matrix{ 0&k&-k\cr -k&0&k\cr k&-k&0}\right)
\ee

\noindent associated  to the following commutations relations among $z_i$

\be 
\matrix{z_1 z_2 & = & w^k z_2 z_1\cr
                 z_1 z_3 & = & w^{-k} z_3 z_1\cr
                 z_2 z_3 & = & w^k z_3 z_2}
\ee

Note that for $k=1$, this algebra has the same structures of the non quartic  K3 studied in \cite{be.le.2} but with $\omega^4=1$. It is simple to see that there are 3-dimensional representations. Indeed, we introduce the two following matrices

\be
P=\left(\matrix{ 1&0&0\cr 0&\omega&0\cr 0&0&\omega^2}\right),\quad
Q=\left(\matrix{ 0&0&1\cr 1&0&0\cr 0&1&0}\right)
\ee

\noindent and so the algebra (\ref{pepita}), for $k=1$,  can be solved  by taking the $z_i$ variables matrices as

\bea
z_1&=&aP\nn\\ 
z_2&=&bQ\\ 
z_3&=&cP^{-1}Q^{-1}\nn
\eea

Note that the  only singularity in the commutative space happens when we take $b=c=0$. The representation theory on the NC algebra becomes reducible at  that point. Therefore, we obtain three distinct irreducible representations.

\item {\bf Case of $n=6$}

Taking $ m=6$, we have the algebraic geometry corresponding  to the  $A_5$ singularity.

\begin{equation}
z_{i }^{6}=x_{i}y_{i},\quad i =1,...,6.
\end{equation}

This equation has $ Z^{5}_6$ discrete symmetry, where  in this case we have  $\sum\limits _{i=1}^6 q^a_i =0, \quad a=1,...,6$. The NC extension (57) is given by the following algebra

\be
z_i z_j=\omega^{\ell_{ij}}z_j z_i,
\ee

\noindent where $\omega$ is a phase such that,  $\omega^6=1$, and $\ell_{ij}$ is $6\times 6$ antisymmetric matrix given by

\be
\ell_{ij}=\pmatrix{ 0                &\ell_{12} &\ell_{13} &\ell_{14} &\ell_{15} & \ell_{16}\cr
                                     -\ell_{12} &        0       &    k_1       &      k_2    &      k_3      &      k_4     \cr 
                                     -\ell_{13} &     -k_1    &       0        &       k_5    &      k_6      &      k_7     \cr
                                     -\ell_{14} &     -k_2    &   -k_5      &         0       &      k_8     &       k_9     \cr
                                     -\ell_{15} &     -k_3    &   -k_6      &      -k_8    &        0        &      k_{10}  \cr
                                     -\ell_{16} &     -k_4    &   -k_7      &      -k_9    &    -k_{10} &        0 }
\ee

\noindent where

\bea
\ell_{12}&=& k_1+ k_2 +k_3 +k_4\nn\\
\ell_{13}&=&-k_1+k_5+k_6+k_7\nn\\
\ell_{14}&=&-k_2-k_5+k_8+k_9\nn\\
\ell_{15}&=&-k_3-k_6-k_8+k_{10}\nn\\
\ell_{16}&=&-k_4-k_7-k_9-k_{10}
 \eea

In what follows we consider the fundamental $6 \times 6$ matrix representation obtained by  using  the following two  matrices set  $Q$, $P_{\eta_{\alpha\beta}};\alpha\beta=1,...,6$ as follows

\be
{P}_{\eta_{\alpha\beta}}={diag(1,\eta_{\alpha\beta},\eta_{\alpha\beta}^{2},...,\eta_{\alpha\beta}^{5})} ,\quad
{Q}=\pmatrix{ 0 & 0 & 0 & 0 & 0 &  1 \cr
                             1 & 0 & 0 & 0 & 0 & 0  \cr 
                             0 & 1 & 0 & 0 & 0 & 0   \cr
                             0 & 0 & 1 & 0 & 0& 0     \cr
                             0 & 0& 0& 1& 0&0         \cr
                            0 & 0 & 0 &0& 1 & 0}
 \ee

\noindent where $\eta _{\alpha\beta}= w^{m_{\alpha\beta}} $ satisfying $\eta_{\alpha\beta}^{6}=1$. From these expressions, it is not difficult to see that the above matrices  satisfy:

\bea
{{P}}_{\alpha }{{P}} _{\beta }&=&{{P}}_{\alpha \beta }\nn\\
{{P}}_{\alpha }^{6}&=&1, \\
 {{Q}}^{6}&=&1. \nn
\eea

Using these  identities and the CY condition,  one can check that the $z_i $  varaibles  can be  presented as

\bea
\label{cosa3}
z_{i}&=&a_{i}\prod_{\alpha,\beta=1}^{6}\left({{P}}_{\eta_{\alpha\beta}}^{q_{i}^{\alpha}}{{Q}}^{q_{i}^{\beta}}\right),\nn\\
x_i&=& b_{i}{\mathbf{I}}_{6}\\ 
y_{i} &=& c_i{\mathbf{I}}_{6} \nn
\eea

In the end of this section  we  would like to  give a comment regarding the reducible representations  for $A_5$ geometry.  We will focus our attention herebelow on giving a particular solution. In this solution we  will consider   an algebra described by ${ Z}_{6}^{2}$ orbifold with ${ Z}_{6}^{3}$ discrete torsions and more  general solutions can be given using similar analysis;  more details can be found in \cite{be.sa3}.  In this way, there exists situations where the representations are reducible. To see this, we start by recalling  that the representation (\ref{cosa3})  corresponds to regular points of NC orbifolds of K3 CY surfaces. These solutions are irreducibles.  However similar solutions may be worked out as well for orbifold points with  the ${ Z}_{6}^{3}$ discrete torsions. Indeed,  choosing matrix coordinates $z_{5}$ and  $z_{6}$ in the centre of the algebra by setting

\be
k_3=k_4=k_6=k_7=k_8=k_9=k_{10}=0,
\ee

\noindent the algebra reduces to

\bea
z_{1}z_{2} &=&w^{k_1+k_2}_{2}z_{2}z_{1}\nn\\ 
z_{1}z_{3}&=&w^{-k_1+k_5}z_{3}z_{1}\nn\\
z_{1}z_{4} &=&w^{-k_2+k_5}z_{4}z_{1}\\
z_{2}z_{3}&=&w^{k_1}z_{3}z_{2}\nn\\
z_{2}z_{4}&=&w^{k_2}z_{4}z_{2},\nn\\ 
z_{3}z_{4}&=&w^{k_5}z_{4}z_{3}\nn
\eea

\noindent and all remaining other relations are commuting. In this equation, the $w$ are such that $w^{6}=1;$ these are the phases of the ${ Z}_{6}^{3}$ discrete torsions. In the singularity where the $z_{1}$, $z_{2},$ $z_{3},$ and $z_{4}$ moduli of equation (\ref{alg1}) act by  zero,  the representation becomes reducible at $z_{1}=z_{2}=z_{3}=z_{4}=0$.

\end{enumerate}

%%%%%%%%%%%%%%%%%%%%%%%%%%%%%%%%%%%%%%%%%%%%%%%%%%%%%%%%%%%%%%%%%%%%%
%%%%%%%%%%%%%%%%%%%%%%%%%%%%%%%%%%%%%%%%%%%%%%%%%%%%%%%%%%%%%%%%%%%%%
\section{Conclusion and Discussions}\label{cd}

In this paper we have studied the NC version of orbifolds of product of K3 surfaces  using the algebraic geometry approach of \cite{be.le.1,be.je.le.1}. In particular we have used a local description of K3 in terms of $A_{n-1}$ geometry to extend the analysis on the NC orbifold torus with discrete torsion initiated in \cite{be.le.1} and exposed explicitly the relation between NC data and the CY charges. Among our results, we have worked out several representations of the corresponding NC algebra by using generic CY charges and given comments regarding the fractionation of branes.

In this context, the ordinary complex deformation of K3  surfaces near an $A_{n-1}$ singularity can be identified with the resolution of stringy singularities by NC algebras using crossed products in the ${\bf C^2}\o {\bf Z_n}$ orbifold space. This  analysis can be generalized to $D$ and $ E$ geometries by replacing the ${\bf Z_n}$ discrete symmetry by the corresponding ones.

On general grounds, it could be said that the appearance of NC geometry when considering discrete torsion is a natural thing. The first appearance of discrete torsion was related to some B-flux on a 2-cycle \cite{vafa.1}, and a relation between the discrete torsion and the torsion part of the homology of the target space was carried in \cite{as.mo.gr}.

The implementation in the presence of D-branes \cite{doug.1,doug.2} makes use of projective representations of the orbifold group, which are classified by $H^2(\Gamma,U(1))$, in perfect correspondence with the previous arguments.

So there is an intimate relation between discrete torsion and the B-field and, in this way, with NC geometry. Even more interesting is the fact that is precisely the presence of this NC geometry which desingularizes the space.

This is important because could be applied to the resolution of singularities not only from a space-time point of view, but in the moduli space of certain theories. For example, a very close case to the ones studied in this paper is that of a $D2$ brane wrapped $n$ times over the fiber of an elliptic $K3$, which can be easily seen to have as moduli space the symmetric product \cite{vafa.2}

\be
{\mathcal{M}}_{1,n}=Sym(K3)=\frac{K3^{\otimes n}}{S_n}
\ee

\noindent where ${\mathcal{M}}_{1,n}$ denotes the moduli space of a D2-brane with charges $(1,n)$ and $S_n$ is the group of permutation of $n$ elements.

On the other hand, the fact that it can be found a reducibility property in the representations of the algebras have lead previously, as we have already mentioned, to an interpretation in terms of the fractionation of branes. However, as we would interpret this configuration as arising as the moduli space of certain configurations, the precise meaning of this result is still not clear for us. However, all these facts will be explored in a future work.

%%%%%%%%%%%%%%%%%%%%%%%%%%%%%%%%%%%%%%%%%%%%%%%%%%%%%%%%%%%%%%%%%%%%%
%%%%%%%%%%%%%%%%%%%%%%%%%%%%%%%%%%%%%%%%%%%%%%%%%%%%%%%%%%%%%%%%%%%%%
\section*{Acknowledgments}
We thank C.~G\'omez, E.H.~Saidi and A.~Uranga for useful conversations and encouragement. A. Belhaj would like to thank Instituto de F\'\i sica Te\'orica, Universidad Aut\'onoma de Madrid for kind hospitality during the preparation of this work. He thanks M. Hssaini  for discussions and for earlier collaboration on this subject. AB work is supported by SARS, programme de soutien \`a la recherche scientifique; Universit\'e Mohammed V-Agdal, Rabat. PR work is supported by an FPU/UAM grant.

%%%%%%%%%%%%%%%%%%%%%%%%%%%%%%%%%%%%%%%%%%%%%%%%%%%%%%%%%%%%%%%%%%%%%
%%%%%%%%%%%%%%%%%%%%%%%%%%%%%%%%%%%%%%%%%%%%%%%%%%%%%%%%%%%%%%%%%%%%%


\begin{thebibliography}{99}


\bibitem{witten.1} E.~Witten,
                                     {\em Noncommutative Geometry and String Field Theory},
                                     Nucl.\ Phys.\ {\bf B268} (1986) 253.


\bibitem{co.do.sc.1} A. ~Connes, M.R. ~Douglas and A. ~Schwarz,
                       {\em Noncommutative Geometry and Matrix Theory: Compactification on Tori},
                       J.\ High Energy Phys.\ {\bf 02} (1998) 003,
                       {\tt hep-th/9711162}.


\bibitem{se.wi.1}N. Seiberg and E. Witten,
                        {\em String theory and Noncommutative Geometry},
                        J.\ High Energy Phys.\ {\bf 09} (1999) 032,
                        {\tt hep-th/990814}.


\bibitem{gr.ne} D. J.~Gross, N.~Nekrasov, 
                              {\em Dynamics of Strings in Noncommutative Gauge Theory}, 
                             J.\ High Energy Phys.\ {\bf 10} (2000) 021,
                             {\tt hep-th/0007204}.


\bibitem{gr.ne.1} D.~Gross and N.~Nekrasov,
                      {\em Monopoles and Strings in Non-commutative Gauge Theory},
                      J.\ High Energy Phys.\ {\bf 07} (2000) 034,
                     {\tt hep-th/0005204}


\bibitem{bak} D.~Bak,
                        {\em Deformed Nahm Equation and a Noncommutative BPS Monopole},
                        Phys.\ Lett.\ {\bf B471} (1999) 149,
                        {\tt hep-th/9910135}.


\bibitem{go.ma} C.~G\'omez and J.J.~Manjar\'\i n,
                                 {\em Dyons, K-theory and M-theory},
                                 {\tt hep-th/0111169}.

\bibitem{ah.be} O.~Aharony and M.~Berkooz, 
                               {\em IR Dynamics of d=2, N=(4,4) Gauge Theories and DLCQ of "Little String Theories''}, 
                              J.\ High Energy Phys.\ {\bf 10} (1999) 030 ,
                              {\tt hep-th/9909101}.

\bibitem{be.sa1}A.~Belhaj and E.H.~Saidi, 
                               {\em Hyperkahler Singularities in Superstrings Compactification and 2-D N=4 Conformal Field Theory},
                                Class.\ Quan.\ Grav. {\bf 18} (2001) 57-82, 
                               {\tt hep-th/0002205}.


\bibitem{be.sa2}A.~ Belhaj and E.H.~Saidi,  
                                {\em On Hyperkahler Singularities},
                               Mod.\ Phys.\ Lett.\ {\bf A15} (2000) 1767-1780,
                               {\tt hep-th/0007143}.



\bibitem{go.mi.st} R.~Gopakumar, S.~Minwalla and A.~Strominger,
                         {\em Noncommutative Solitons},
                         J.\ High Energy Phys.\ {\bf 05} (2000) 020,
                        {\tt hep-th/0003160}.


\bibitem{sa.sa} E.M.~Sahraoui and E.H.~Saidi,
                          {\em Solitons on Compact and Noncompact Spaces in Large Noncommutativity},
                         Class.\ Quan.\ Grav.\ {\bf 18} (2001) 3339-3358.
                         {\tt hep-th/0012259}.


\bibitem{be.be.di.sa} I.~Benkaddour, M.~Bennai, E.~Diaf and H.~Saidi
                         {\em  On Matrix Model Compactification on Non-commutative $F_0$ Geometry},
                         Class.\ Quan.\ Grav.\ {\bf 17} (2000) 1765.


\bibitem{be.le.1} D.~Berenstein and R.G.~Leigh,
                                {\em Non-Commutative Calabi-Yau Manifolds},
                                Phys.\ Lett.\ {\bf B499} (2001) 207-214,
                                {\tt hep-th/0009209}.

\bibitem{be.sa3} A.~Belhaj and  E.H.~Saidi,  
                                 {\em On Non Commutative Calabi-Yau Hypersurfaces}, 
                                 Phys.\ Lett.\ {\bf B523} (2001) 191-198,
                                 {\tt hep-th/0108143}.


\bibitem{lo.sc.we} A.~Lorek, W.B.~Schmidke and J.~Wess, 
                                   {\em $SU_q(2)$ Covariant R Matrices for Reducible Representations}, 
                                  Lett.\ Math.\ Phys.\ {\bf 31} (1994) 279.


\bibitem{we} J.~Wess, 
                          {\em q-Deformed Heisenberg Algebras}, 
                          {\tt math-ph/9910013}.


 \bibitem{ki.lr1} H.~Kim and C-Y.~Lee, 
                                {\em Noncommutative K3 Surfaces}, 
                                Phys.\ Lett.\ {\bf B536} (2002) 154-160,
                                {\tt  hep-th/0105265}.


\bibitem{gr.va.wa.1} B.R.~Greene, C.~Vafa and N.P.~Warner,
                                        {\em Calabi-Yau Manifolds and Renormalization Group Flows},
                                        Nucl.\ Phys.\ {\bf B324} (1989) 371.

\bibitem{va.wi.1} C.~Vafa and E.~Witten,
                                  {\em On Orbifolds with Discrete Torsion},
                                  J.\ Geom.\ Phys.\ {\bf 15} (1995) 189-214,
                                  {\tt hep-th/9409188}.

\bibitem{ka.ma.va}S.~Katz, P.~Mayr and C.~Vafa,
                                     {\em Mirror Symmetry and Exact solution of 4-D N=2 Gauge Theories: 1},
                                     Adv.\ Theor.\ Math.\ Phys.\ {\bf 1} (1998) 53.
                                     {\tt hep-th/9706110}

\bibitem{ma} P.~Mayr, 
                          {\em N=2 Gauge Theories},
                         Lectures presented at Spring school on superstring theories and related matters, ICTP, Trieste, Italy, (1999).


\bibitem{be.fa.sa} A.~Belhaj, A.E.~Fallah and E.H.~Saidi, 
                                  {\em On the Non-simply Laced Mirror Geometries in Type II Strings},
                                  Class.\ Quan.\ Grav.\ {\bf 17} (2000) 515-532.


 \bibitem{be.le.2} D.~Berenstein and R.G.~Leigh,
                                {\em Resolution of Stringy Singularities by Non-commutative Algebras},
                                J.\ High Energy Phys.\ {\bf 06} (2001) 030,
                                {\tt hep-th/0105229}.


\bibitem{be.je.le.1} D.~Berenstein, V.~Jejjala and R.G.~Leigh,
                         {\em Marginal and Relevant Deformations of N=4 Field Theories and Non-Commutative Moduli Spaces of Vacua},
                         Nucl.\ Phys.\ {\bf B589} (2000), 196-248,
                         {\tt hep-th/0005087}.


\bibitem{sa}E. H.~Saidi, 
                       {\em NC Geometry and Discrete Torsion Fractional Branes:I},
                       {\tt hep-th/0202104}.


\bibitem{ki.lr2} H.~Kim and C-Y.~Lee,
                               {\em N-Point Deformation of Algebraic K3 Surfaces}, 
                               {\tt  hep-th/0204013}.


\bibitem{vafa.1} C.~Vafa,
                              {\em Modular Invariance and Discrete Torsion on Orbifolds},
                              Nucl.\ Phys.\ {\bf B273} (1986) 592. 

\bibitem{as.mo.gr} P.S.~Aspinwall, D.R.~Morrison and M.~Gross,
                                     {\em Stable Singularities in String Theory},
                                     Comm.\ Math.\ Phys.\ {\bf 178} (1996) 115-134,
                                     {\tt hep-th/9503208}.

\bibitem{doug.1} M.R.~Douglas,
                                  {\em D-branes and Discrete Torsion},
                                  {\tt hep-th/9807235}.

\bibitem{doug.2} M.R.~Douglas and B.~Fiol,
                                 {\em D-branes and Discrete Torsion. II},
                                 {\tt hep-th/9903031}.

\bibitem{vafa.2} C.~Vafa,
                                 {\em Lectures on Strings and Dualities},
                                Published in ``Trieste 1996, High energy physics and cosmology''
                                 {\tt hep-th/9702201}.

\end{thebibliography}
\end{document}